\documentclass[journal=jacsat,manuscript=article]{achemso}

\usepackage{filecontents}

\usepackage{amsmath}
\usepackage{amssymb}
\usepackage{graphicx}
\usepackage[colorlinks=true, allcolors=blue]{hyperref}
\usepackage{booktabs}
\usepackage{multirow}
\usepackage[version=4]{mhchem} 
\usepackage{subcaption}
\usepackage{tablefootnote}
\usepackage{siunitx} 
\usepackage{xr-hyper}
\def\ang{$\text{\AA}$}

\author{Chikashi Shinagawa}
\email{shinagawa@preferred.jp}
\author{So Takamoto}
\author{Daiki Shintani}
\author{Yong-Bin Zhuang}
\author{Yuta Tsuboi}
\author{Katsuhiko Nishimra}
\author{Kohei Shinohara}
\author{Shigeru Iwase}
\affiliation{Preferred Networks, Inc., Otemachi Bldg. 1-6-1 Otemachi, Chiyoda-ku, Tokyo, 100-0004, Japan}
\author{Yuta Tanaka}
\affiliation{AI Innovation Department, ENEOS Holdings, Inc., 8, Chidoricho, Naka-ku, Yokohama, Kanagawa 231-0815, Japan}
\author{Ju Li}
\affiliation{Department of Nuclear Science and Engineering and Department of Materials Science and Engineering, Massachusetts Institute of Technology, Cambridge, MA, 02139, USA}

\title{Matlantis-PFP v8: Universal Machine Learning Interatomic Potential with Better Experimental Agreements via r$^2$SCAN Functional}

\abbreviations{IR,NMR,UV}

\begin{document}

\begin{abstract}
 
Universal Machine Learning Interatomic Potentials (uMLIPs) enable atomistic simulations and high-throughput screening at scales far beyond those accessible with density functional theory (DFT).  However, most existing uMLIPs are trained on Perdew--Burke--Ernzerhof (PBE) generalized gradient approximation (GGA) data and are therefore fundamentally limited by PBE-level accuracy.  In this paper, we argue that better zero-shot predictions versus experiments must be an explicit design target for uMLIPs and present PFP v8, a uMLIP available on the Matlantis service that overcomes the inherent limitations of the PBE functional by being trained to reproduce the regularized-restored strongly constrained and appropriately normed (r$^2$SCAN) meta-GGA potential-energy surface across a wide range of chemical domains. Without requiring domain-specific fine-tuning, PFP v8 delivers systematically improved agreement with experimental data or high-accuracy references for crystals, molecules, and surfaces, outperforming PBE-based DFT calculations. Crucially, in long-time molecular dynamics simulations that are computationally impractical with DFT, PFP v8 predicts melting points with an average error of approximately 130 K, halving the error relative to PBE-trained models. These results establish that uMLIPs can move beyond the limitations of their training approximations and achieve substantially improved agreement with experiment across diverse chemical domains, further narrowing the gap between simulation and reality.

\end{abstract}

\section{Introduction}
Machine learning interatomic potentials (MLIPs) have become powerful tools in materials science. By drastically reducing computational costs while maintaining density functional theory (DFT) level accuracy, MLIPs enable simulations of larger sizes and longer time scales, while also making it practical to systematically explore diverse structural candidates, thereby bridging the gap between reality and simulation. After early successes in modeling organic molecules composed of a few elements~\cite{ani-1,ani-2}, MLIPs have evolved to handle diverse elements~\cite{oc20,m3gnet}, and currently, numerous MLIPs cover most elements of the periodic table.~\cite{chgnet,nequip,mace-mpa0} While MLIPs were initially evaluated on specific limited tasks, such as adsorption energies~\cite{oc20} or crystal stability~\cite{matbenchdiscovery}, the focus has recently shifted toward universal MLIPs (uMLIPs) applicable to diverse domains, leading to proposals for more comprehensive evaluation metrics~\cite{mace-mpa0,chips-ff} and research on uMLIPs trained on multiple datasets with different domains and computational conditions~\cite{uma,sevennet-omni}.

Most existing uMLIPs or MLIPs capable of handling numerous elements~\cite{m3gnet,chgnet,mace-mpa0,GNoME,MatterSim,EquiformerV2,SevenNet,eSEN,FlashTP,PET-MAD,Orb,Orb-v3,TACE,GACE-OAM,DPA-3} are trained on datasets generated via DFT calculations using the Perdew--Burke--Ernzerhof (PBE) generalized gradient approximation (GGA) exchange-correlation (XC) functional~\cite{GGA-PBE}, and are thus expected to reproduce the potential energy surface (PES) of the PBE functional. While the PBE functional has been widely used in materials research for decades as the standard for DFT calculations (excluding isolated structures like molecules), it is well known to have issues in the comparison with experimental values. For example, regarding formation energy, which is one of the most fundamental properties of crystal structures, DFT calculations using the PBE functional (DFT-PBE) are known to exhibit errors of 50--200 meV/atom relative to experimental values~\cite{Kingsbury2022}. The mean absolute error (MAE) of the latest MLIPs for crystal formation energy relative to DFT-PBE is tens of meV/atom~\cite{matbenchdiscovery}, which is smaller than the intrinsic error of PBE itself. In other words, there is a concern that further reducing the regression error of MLIPs against the PBE PES may not necessarily improve their ability to compare with experimental observations, which is the fundamental objective of MLIPs. Consequently, a paradigm shift is required: from merely emulating a specific DFT functional to explicitly targeting experimental comparison.

With the aim of accelerating materials discovery and research for diverse domains and unknown materials, we reported PreFerred Potential (PFP) as a ``towards universal MLIP''~\cite{takamoto2022towards}. It is capable of reproducing diverse domains---including crystals, molecules, surfaces, and adsorption structures---for 45 elements without the need for fine-tuning (i.e. off-the-shelf). PFP is provided through Matlantis (\url{https://matlantis.com}), a Software-as-a-Service platform, and has been applied to batteries~\cite{pfp-usecase-battery:10.1021/jacs.5c00856,pfp-usecase-battery:10.1002/adma.202210055,pfp-usecase-battery:10.1002/aenm.202304085,pfp-usecase-battery:10.1021/acs.chemmater.4c02852,pfp-usecase-battery:10.1021/acs.jpcc.3c02379}, catalysts~\cite{pfp-usecase-catalyst:10.1021/jacsau.3c00018,pfp-usecase-catalyst:10.1039/D4SC04790G,pfp-usecase-catalyst:10.1038/s41929-024-01236-y,pfp-usecase-catalyst:10.1016/j.jcat.2025.116253,pfp-usecase-catalyst:10.1021/acs.jpcc.3c08093,pfp-usecase-catalyst:10.1021/acs.jpcc.4c01831,pfp-usecase-catalyst:10.1038/s41467-025-66242-5}, semiconductors~\cite{pfp-usecase-semiconductor:10.1002/advs.202516784,pfp-usecase-semiconductor:10.1126/sciadv.adm7221,pfp-usecase-semiconductor:10.1039/D5MA00220,pfp-usecase-semiconductor:10.1002/smll.202402543} and other materials or domains~\cite{pfp-usecase-glass:10.1111/jace.19629,pfp-usecase-other:10.1021/jacs.4c09476,pfp-usecase-other:10.1021/acsnano.4c13792,pfp-usecase-other:10.1021/acscatal.4c07737,pfp-usecase-other:10.1002/advs.202307417,pfp-usecase-other:10.1016/j.jnucmat.2026.156478,pfp-usecase-other:doi:10.1021/acs.chemmater.4c03298} without fine-tuning. PFP v7, released on Matlantis in September 2024, supports 96 elements,~\cite{JACOBS2025101214} covering all naturally occurring elements. By combining this extensive elemental coverage with the inherent capability to simulate a wide range of chemical environments and phenomena, PFP has advanced toward the realization of a universal MLIP. However, like most existing MLIPs, PFP was trained to learn and reproduce the PES of the PBE functional, meaning that its predictive power was capped by the inherent limitations of the PBE functional.

One promising route to further bridging the gap between reality and simulation beyond these limitations is to climb ``Jacob's ladder''~\cite{perdew2001jacob} in DFT and train MLIPs on the PES of XC functionals at a higher level than GGA. For example, OMol25~\cite{omol25} is a large-scale dataset of molecules and complexes with 83 atoms, constructed using the range-separated hybrid functional $\omega$B97M-V~\cite{wb97MV}, which resides two rungs above GGA on Jacob’s ladder. While $\omega$B97M-V has demonstrated high accuracy for such molecular domains, it contains a range-separation parameter ($\omega$), for which the optimal choice can be domain- and material-dependent. For example, HSE range-separated hybrid functionals~\cite{hse03,hse06}, often used for semiconductors, incorporate short-range Hartree–Fock (HF) exchange, while $\omega$B97-type functionals~\cite{wB97}, including $\omega$B97M-V, primarily incorporate long-range HF exchange. This domain-dependent character of range-separated hybrids complicates the establishment of a single, consistent reference level suitable for universal MLIPs applicable to diverse domains. Moreover, even for global hybrid functionals without a range-separation parameter such as PBE0~\cite{pbe0}, the explicit evaluation of HF exchange leads to a higher computational complexity order than the GGA functional, making it challenging to construct datasets, especially for periodic structures.

The regularized-restored strongly constrained and appropriately normed (r$^2$SCAN) functional~\cite{r2scan} is a meta-GGA functional, residing one rung above GGA. Although range-separated hybrid functionals can provide higher accuracy for specific domains, the r$^2$SCAN functional has been reported to systematically improve agreement with experimental observations across a wide range of domains including molecules~\cite{ehlert_r2scan-d4_2021} and various crystals~\cite{Kingsbury2022,acsmaterialsau.2c00059} without any tuning parameters. Importantly, the r$^2$SCAN functional retains the same formal computational scaling as the GGA functional and exhibits favorable numerical convergence; its practical computational cost is only about four times that of PBE~\cite{PhysRevB.102.121109}. These desirable characteristics make r$^2$SCAN the optimal XC functional for uMLIPs at present, and it has already been adopted by the Materials Project and has also been used to construct MLIP datasets such as MatPES~\cite{matpes} and MP-ALOE~\cite{mp-aloe}. However, these efforts are limited to crystals and their vicinities. Given the recognized importance of datasets generated under identical computational conditions across diverse domains~\cite{sevennet-omni}, realizing a universal r$^2$SCAN-based MLIP calls for a dataset that is not restricted to crystals and their vicinities, but instead spans a broad range of domains computed consistently with the r$^2$SCAN functional.

In this study, by employing the r$^2$SCAN functional, we developed PFP v8, a uMLIP capable of providing better accuracy when compared against experimental values for diverse domains across 70 elements. We prioritized universality across diverse domains over elemental coverage to ensure rigorous experimental reproducibility, deferring the inclusion of lanthanides and heavy elements to the next version. PFP v8 was trained on an originally constructed dataset containing 3 million structures using the r$^2$SCAN functional, combined with our previously constructed dataset containing 60 million structures using the PBE functional. We compared the results of PFP v8 against experimental values or high-precision calculation results achieving chemical accuracy for crystals, molecules, and surfaces. The results demonstrate that PFP v8 outperforms PBE-based baselines across all domains, notably achieving a formation energy error of 80 meV/atom for crystals and predicting surface energies comparable to experimental uncertainty. Furthermore, we used PFP v8 to calculate melting points via long-duration molecular dynamics (MD) simulations---which are practically impossible with standard DFT calculations due to computational costs---and compared the results with experimental values. These results demonstrate that PFP v8, employing the r$^2$SCAN functional, exhibits experimental reproducibility that significantly surpasses that of the PBE functional, indicating that PFP v8 has further bridged the gap between reality and simulation.

\section{Methods}
\label{sec:methods}
\subsection{Dataset domains and structures}
In our previous work~\cite{takamoto2022towards}, we reported the PFP-WB97XD dataset (previously referred to as the PFP molecular dataset) based on the $\omega$B97X-D functional~\cite{wB97XD} for molecules, and the PFP-PBE+U dataset (previously referred to as the PFP crystal dataset) based on the PBE functional with Hubbard $U$ correction~\cite{HubbardU}, which included various structures. We have since extended the PFP-PBE+U dataset to cover 96 elements~\cite{JACOBS2025101214} and also constructed the PFP-PBE dataset based on the PBE functional without the Hubbard $U$ correction.

\begin{figure}[htb]
    \centering
    \includegraphics[width=\linewidth]{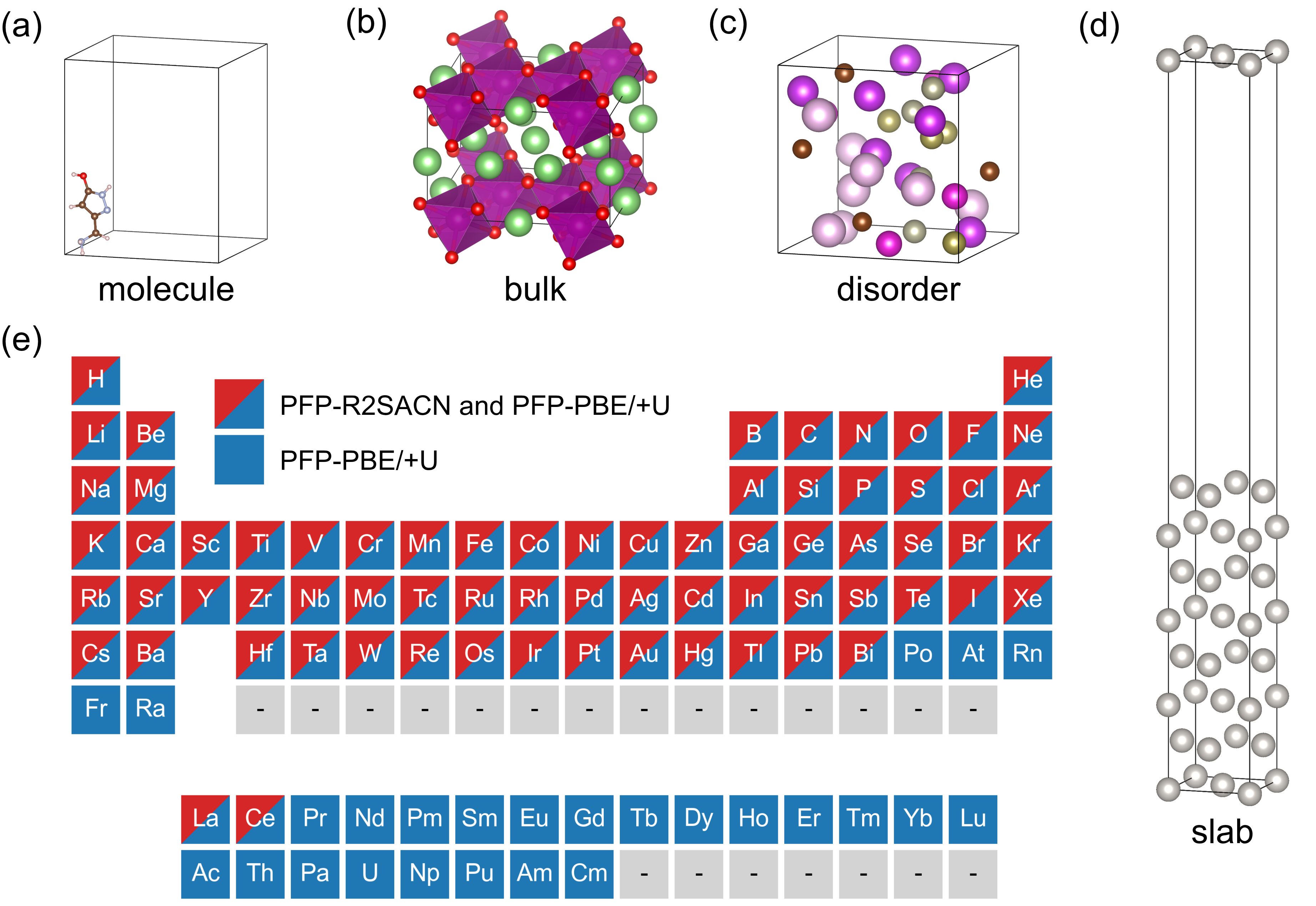}
    \caption{Overview of our original datasets. Examples of structures included in the PFP-R2SCAN dataset: (a) molecule, (b) bulk, (c) disorder, and (d) slab structures, visualized by VESTA~\cite{momma2011vesta}. (e) Elemental coverage of our original datasets. Elements highlighted in both red and blue are included in both the PFP-R2SCAN and PFP-PBE/+U datasets (70 elements). Elements highlighted in blue are included only in the PFP-PBE/+U dataset.}
    \label{fig:datasets}
\end{figure}

We have now newly constructed our original PFP-R2SCAN dataset using the r$^2$SCAN functional. The domains, number of elements, and data points included in these datasets, along with the existing ones, are summarized in Table~\ref{tab:datasets}. We believe that a dataset containing diverse structures is indispensable for a uMLIP. As shown in Table~\ref{tab:datasets} and Figure~\ref{fig:datasets}, the PFP-R2SCAN dataset contains diverse domains, such as molecules, bulk, slab, and disordered structures, and covers 70 elements. For typical structures and the preparation methods of these domains, please refer to the previous work~\cite{takamoto2022towards}. The ``disorder'' domain represents the most distinctive structural feature of our dataset. These structures were generated by randomly selecting up to 20 elements and simulating them at high temperatures using PFP-based MD. While these structures significantly deviate from practical simulation scenarios, they are expected to improve predictions across diverse chemical domains~\cite{takamoto2022neurips}.

Although the covered domains are more limited compared to the PFP-PBE or PFP-PBE+U datasets (hereafter referred to as the PFP-PBE/+U dataset), the PFP-R2SCAN dataset still covers a variety of domains under identical computational conditions when compared to other existing datasets developed for specific tasks, such as molecules, bulk or adsorption structures. Due to the fact that r$^2$SCAN functional calculations take approximately 4 times longer than PBE functional calculations, and because we spent only about one year constructing the PFP-R2SCAN dataset compared to approximately five years for the PFP-PBE/+U dataset, the size of the new dataset is only about 1/20th that of the PFP-PBE/+U dataset.

\begin{table*}[h]
    \centering
    \caption{Summary of our original datasets and public dataset used for PFP v8 training}
    \label{tab:datasets}
    \resizebox{\textwidth}{!}{
    \begin{tabular}{l|c|cccccc|cc}
        \toprule
        & & \multicolumn{6}{c|}{domains} & \multicolumn{2}{c}{\# of} \\
        \midrule
        Dataset & XC functional & molecule & bulk & slab & disorder & cluster & adsorption & Elements & Data \\
        \midrule
        PFP-R2SCAN & r$^2$SCAN & \checkmark & \checkmark & \checkmark & \checkmark & & & 70 & $3\times10^6$ \\
        PFP-PBE/+U & PBE & \checkmark & \checkmark& \checkmark& \checkmark& \checkmark& \checkmark & 96 & $6\times10^7$ \\
        PFP-WB97XD & $\omega$B97X-D & \checkmark & & & & & & 9 & $6\times10^6$ \\
        OC20~\cite{oc20} & PBE & & & & & & \checkmark & 56 & $1\times10^8$ \\
        \bottomrule
    \end{tabular}
    }
\end{table*}

\subsection{DFT calculation}
The original r$^2$SCAN dataset was constructed with collinear spin-polarized DFT calculations using the Vienna ab-initio simulation package (VASP)~\cite{vasp,vasp_paw} version 6.4.3 with GPU acceleration implemented via OpenACC. The r$^2$SCAN meta-GGA functional~\cite{r2scan} was employed to approximate the exchange-correlation energy. The projector-augmented wave (PAW) method~\cite{blochl1994paw,vasp_paw} was utilized to describe the interaction between core and valence electrons, using the latest standard PBE (PBE\_64) PAW potentials provided with VASP. While the PFP-PBE/PBE+U dataset covers 96 elements, the current PFP-R2SCAN dataset focuses on 70 elements. While the selection of PAW potentials for f-block elements presents challenges even in PBE calculations, we temporarily excluded most of lanthanides because the optimal selection of PAW potentials and valence configurations for f-block elements under the r$^2$SCAN functional requires further rigorous validation to match our standards for experimental reproducibility. Additionally, elements from Po to Cm were excluded primarily due to their relatively lower priority in our current target applications. Extension to these elements is planned for the next version (PFP v9). To reproduce the PES in the ground state, we performed calculations using initial magnetic moments corresponding to multiple magnetic orderings, including ferromagnetism and antiferromagnetism, and adopted the results showing the lowest energy. However, for some systems, we conducted calculations exclusively under ferromagnetic conditions. Hubbard $U$ correction was not applied.

To realize a uMLIP with high accuracy vs. experimental values across a wide range of domains based on the r$^2$SCAN functional, we determined detailed computational conditions considering both universality and accuracy. Following existing research~\cite{PhysRevMaterials.6.013801}, we adopted an energy cutoff of 680 eV for the plane-wave basis, which is higher than that used for PFP-PBE+U. Considering universality, we adopted a method to determine the number of $k$-points based on the length in reciprocal space (KSPACING = 0.22~\cite{PhysRevMaterials.6.013801}, corresponding to one $k$-point approximately every $1/28.56~\ang^{-1}$), replacing the previous method used for PFP-PBE+U which also considered the number of atoms (1000 $k$-points per reciprocal atom). For instance, in slab structure calculations for determining surface energy, the number of in-plane $k$-points is set to be identical to that of the bulk structure to expect error cancellation in $k$-point convergence, while only the $\Gamma$-point is sampled in the out-of-plane direction. Although the previous method considering the number of atoms cannot satisfy this requirement, the method adopted here naturally does so, provided a sufficiently thick vacuum layer is set. Following verification of total energy convergence, the vacuum layer thickness was set to 20 \ang~for slab structures and 10 \ang~in each direction for molecular structures, ensuring that only the $\Gamma$-point is sampled in the direction of any vacuum layer. To maintain consistency between energy and force, Gaussian smearing with a width of 0.05 eV was used to determine the partial occupancies of each orbital, as in previous work. Furthermore, more detailed computational conditions were determined by referring to the benchmark report~\citet{Kaplan_Pymatgen_Issue_3322} by the authors of MatPES~\cite{matpes}. These conditions are summarized in the Supporting Information section~\ref{si:dft_details} and are nearly identical to the computational conditions used for MatPES.

The computational conditions used for the PBE+U dataset are detailed in our previous work~\cite{takamoto2022towards} and the Supporting Information section~\ref{si:dft_details}. Furthermore, the computational conditions for the PBE dataset are identical to those used for the PBE+U dataset, except that the Hubbard $U$ correction was not applied.

\subsection{MLIP Architecture}
Recent progress of the MLIP development is driven by message-passing graph neural networks, which represent atomic systems as graphs and learn environment-dependent many-body interactions~\cite{schnet,dimenet,dimenetplusplus,painn,gemnet,TAKAMOTO2022111280,nequip,mace}. By construction, MLIPs typically respect key physical symmetries such as translational, rotational, and permutational equivariance.
The architecture of PFP v8 is based on the Tensor Embedded Atom Network (TeaNet) \cite{TAKAMOTO2022111280}, a graph neural network (GNN) framework originally introduced in the initial version of PFP. A key characteristic of TeaNet is its incorporation of higher-order Euclidean tensors (rank-2 tensors) into the message-passing mechanism, in addition to conventional scalars and vectors. These rank-2 tensors were specifically proposed as essential information to reformulate the interactions of bond-order type interatomic potentials into a graph convolution framework. The introduction of rank-2 tensors allows the model to naturally represent and propagate higher-order geometric features, such as bond angles and dihedral angles without explicitly introducing the three or four-body interaction terms which cause the combinatorial explosion.

By stacking multiple graph convolution layers, the architecture facilitates message passing that enables interactions between atoms over longer distances. This mechanism is expected to capture complex, angle-dependent multi-body effects, including dihedral angles and $\pi$-bonding interactions. The internal operations of the graph convolutions consist of inner products and tensor products, ensuring that the architecture is equivariant under rotation and reflection operations. Consequently, the final scalar output, the total energy, is naturally invariant with respect to these operations. Atomic forces and stresses are derived as the derivatives of the energy with respect to atomic coordinates and cell shape, respectively. To enhance the physical descriptiveness of the model, Bader charges (scalar values per atom) are incorporated as an auxiliary modality during the training process. The design of TeaNet is physically inspired by the iterative process of electronic relaxation. Each of its layers can be interpreted as a single ``timestep" in emulating the propagation of orientation-dependent electronic structure information throughout the atomic network. Previous numerical experiments have confirmed that both the inclusion of higher-order tensors and the increased number of stacked convolution layers contribute to architectural performance enhancements \cite{TAKAMOTO2022111280}.

For input, the model takes a list of atomic species, coordinates, and cell parameters alongside a specific ``calculation mode'' (as detailed in the next Section). For the output value (the energy), it is designed to be arbitrarily differentiable with respect to input coordinates. In PFP v8, the maximum cutoff distance for the last two graph convolution layers was increased from $0.6$ nanometers to $0.9$ nanometers, allowing a more comprehensive description of long-range environments.

From an implementation standpoint, the model is trained using PyTorch with CUDA environments. All inference computations are performed on NVIDIA GPUs using the Matlantis-optimized mode of PFVM, our proprietary inference engine.

\subsection{Training with multiple datasets} \label{sec:calc_mode}
A unique challenge in developing a universal model is reconciling multiple datasets generated under inconsistent DFT conditions, such as different exchange-correlation functionals (e.g., PBE vs. $\omega$B97X-D) or basis sets. Directly merging these datasets would typically introduce unintended energy gaps and harm model performance. PFP addresses this through its ``Calculation Mode'' mechanism, where labels corresponding to specific DFT conditions are assigned to the data during training. This multi-dataset training strategy has been a core feature of PFP since its inception in 2022 \cite{takamoto2022towards}. It is noted that the method of incorporating multiple datasets has become a major research topic in recent years \cite{uma,sevennet-omni}. This allows the single model to concurrently learn multiple, mutually contradictory energy surfaces with high accuracy. During inference, users can specify the desired calculation mode to select the appropriate DFT condition. This approach also facilitates domain transfer, where properties originally computed under one DFT condition can be effectively transferred to enhance inference under another condition. For the architectural treatment, given that the TeaNet architecture does not utilize a global node, these labels are embedded into the features of individual nodes. While it is architecturally possible to implement this behavior through specialized output layers or by conditioning the neural network parameters on the labels, we have opted for the current node-based input approach for its simplicity and its intuitive interpretation as a flavor assigned to each element information.

The robustness of this approach is demonstrated by the model's ability to generalize across disparate chemical domains. For instance, in PFP v0, the $\omega$B97X-D dataset primarily contained light organic elements such as H, C, N, and O, and lacked metallic elements such as Li or Na. However, by training simultaneously on a PBE+U dataset that included these metals, the model produced physically reasonable outputs even when the calculation mode was set to $\omega$B97X-D for structures containing metallic species. Numerical experiments confirm that in regions where mode-specific data is abundant (e.g., organic molecules in $\omega$B97X-D mode), the model accurately reproduces structural parameters for that functional. Conversely, in regions where data for a specific mode is missing, the model performs informed inference by extrapolating from other available calculation modes. This capability allows MLIPs to maintain structural and energetic consistency across a vast chemical space that no single functional-specific dataset could fully cover.

The computational modes corresponding to each dataset shown in Table \ref{tab:datasets} are provided in PFP v8, as summarized at the top of Table \ref{tab:calc_mode}. Note that although computational modes exist for the OC20 dataset, they are not provided in the Matlantis service.

\subsection{DFT-D3 Correction}
\label{sec:dft-d3}
Long-range interactions, such as van der Waals (vdW) forces, are essential in chemical simulations; therefore, these effects must be explicitly addressed in simulations utilizing MLIPs. Although it is technically possible to train a model on datasets that include long-range terms, such an approach is fundamentally not suited to our PFP architecture, which relies on local cutoffs. To resolve this, we adopted a strategy in which the dispersion interaction is treated separately. Since the DFT-D3 correction~\cite{grimme_consistent_2010} does not require electronic structure information, we calculate this term independently and superimpose it onto the PFP prediction.

For each calculation mode without DFT-D3 correction excluding PFP-WB97XD, we provided the corresponding calculation modes with D3 correction as shown in Table \ref{tab:calc_mode}.
Although the original DFT-D3 correction is fast compared to DFT calculations, it is comparable in speed to MLIP inference.
To avoid degradation in the overall inference speed, we utilized the faster torch-dftd package~\cite{takamoto2022towards,TorchDFTD} for DFT-D3 corrections.
Additionally, while the original DFT-D3 correction uses a cutoff distance of 95 Bohr~\cite{Grimme-D3-Manual}, we have shortened it to 14 \ang~in order to improve the inference speed and the number of atoms that can be handled without losing accuracy.
We used the Becke--Johnson damping function and the recommended parameters for PBE~\cite{grimme_effect_2011} and r$^2$SCAN~\cite{ehlert_r2scan-d4_2021}, respectively. We did not provide +D3 modes for Am and Cm because there are no suitable parameters available.

\begin{table*}[h]
    \centering
    \caption{Summary of calculation modes (calc\_mode) of PFP v8 about XC functional, Hubbard $U$ correction, DFT-D3 correction, and the number of supported elements.} 
    \label{tab:calc_mode}
    \begin{tabular}{lcccc}
         \toprule
         calc\_mode & XC functional & \begin{tabular}[c]{@{}c@{}}Hubbard $U$ \\ Correction \end{tabular} & \begin{tabular}[c]{@{}c@{}} DFT-D3 \\ Correction \end{tabular}& \begin{tabular}[c]{@{}c@{}} \# of \\
         Supported Elements\end{tabular}\\
         \midrule
         R2SCAN & r$^2$SCAN &  &  & 70 \\
         PBE & PBE & & & 96 \\
         PBE+U & PBE & \checkmark & & 96 \\
         WB97XD & $\omega$B97X-D & & & 9 \\
         \midrule
         R2SCAN+D3 & r$^2$SCAN &  & \checkmark & 70 \\
         PBE+D3 & PBE & & \checkmark & 94 \\
         PBE+U+D3 & PBE & \checkmark & \checkmark & 94 \\
         \bottomrule
         \end{tabular}
\end{table*}

\section{Results and Discussion}

\subsection{Validation of r$^2$SCAN-level accuracy across diverse chemical domains}
\paragraph{\label{sec:results_formation_enthalpy}Crystal formation energies}

We evaluated the formation energies of crystal structures with PFP and subsequently compared them with experimental standard formation enthalpies, which were retrieved from the \texttt{expt\_formation\_enthalpy\_kingsbury} dataset~\cite{wang_framework_2021} distributed with Matminer~\cite{6ddf244882c24092addcff4be1eb7ce1}.
The formation energy data were compiled primarily from the Kubaschewski tables \cite{Kubaschewski1993}, the NIST JANAF database \cite{chase_nist-janaf_1998}, and the compilation of Kim et al. \cite{kim_experimental_2017}.

For the above dataset, we performed structural optimizations including cell volume and shapes, and estimated formation energies using PFP-R2SCAN and PFP-PBE/+U.
For PFP-PBE, we considered both cases with and without correction.
In the corrected case, we switched between using the PBE and PBE+U based on the system and applied a correction for the GGA/GGA+U mixing scheme~\cite{wang_framework_2021,jain_formation_2011} in the Materials Project.
For reference, the DFT-r$^2$SCAN formation energies were retrieved from the Materials Project~\cite{jain_commentary_2013,horton_accelerated_2025}.

Table~\ref{tab:formation_energy} summarizes the results from each calculation mode and the reference DFT data of Materials Project in comparison with experimental values.
In this comparative analysis, we restricted the target structures to those containing only the 70 elements supported by PFP-R2SCAN and for which reference data exists in the Materials Project using the r$^2$SCAN functional, totaling 1,100 structures.
Note that the formation energy is compared to the standard formation enthalpy without considering the temperature, pressure, and zero-point vibration.

\begin{table*}[ht]
\centering
\caption{MAE of formation energies with each calculation mode of PFP and the DFT results from Materials Project~\cite{jain_commentary_2013,horton_accelerated_2025} compared to the experimental values~\cite{wang_framework_2021}.}
\label{tab:formation_energy}
\begin{tabular}{lcccc}
\toprule
& PFP-R2SCAN & PFP-PBE/+U & PFP-PBE & DFT-r$^2$SCAN \\
\midrule
MAE (eV/atom) & 0.080 & 0.121 & 0.162 & 0.080 \\
\bottomrule
\end{tabular}
\end{table*}

Predictions using PFP-PBE/+U with corrections showed improved accuracy compared to PFP-PBE alone. 
However, the MAE was further reduced by using PFP-R2SCAN without any corrections.
Notably, the MAE of PFP-R2SCAN is comparable to that of the reference DFT-r$^2$SCAN results from Materials Project.
These results demonstrate the superiority of PFP-R2SCAN, trained on high-precision r$^2$SCAN data, in predicting crystal-structure stability.

\paragraph{Molecular benchmarks: GMTKN55}
\label{sec:gmtkn55}
To rigorously assess the performance of our PFP model for molecules, we evaluated it against the GMTKN55 dataset~\cite{C7CP04913G}. GMTKN55 is a comprehensive benchmark database designed to test the accuracy of quantum chemical methods across a wide range of chemical applications. It comprises 55 subsets containing a total of 1,505 relative energies, covering barrier heights, reaction energies, and both intramolecular and intermolecular noncovalent interactions. This dataset provides reference values of significantly higher quality for most sets, often relying on highly accurate composite protocols like the Weizmann methods.

Due to inherent limitations in the current PFP framework, the original GMTKN55 dataset was preprocessed for our evaluation. PFP cannot handle structures with a net charge, so all charged molecules were excluded from the benchmark subsets.  Additionally, the isolated hydrogen (H) atom was excluded because its reference energy was incompatible with the PFP model. After this preprocessing, 1,263 out of the original 1,505 relative energies were used for the evaluation. The subset names used in the experiment are shown in the Supporting Information section~\ref{si:gmtkn55}.

The primary metric for this evaluation is the weighted total mean absolute deviation, known as WTMAD-2. This metric provides a balanced assessment by applying a weighting factor to the mean absolute deviation (MAD) of each subset. The weighting is based on the difficulty of the subset, which is inversely proportional to the average absolute reference energy of the reactions within it. For consistent comparison, all DFT results from the literature were re-evaluated using the same preprocessed dataset, recalculating their WTMAD-2 scores based only on the 1,263 data points employed in the PFP evaluation.

\begin{table*}[ht]
\centering
\caption{WTMAD-2 scores for different calculation modes (calc\_mode) of PFP and DFT functionals on the GMTKN55 benchmark subsets (kcal/mol). The Becke–Johnson damping function is also employed in DFT-D3 corrections in DFT calculations.}
\label{tab:gmtkn55_results}
\begin{tabular}{llccccccc}
\toprule
Method & \begin{tabular}[c]{@{}c@{}}calc\_mode \\or functional\\(+ dispersion)\end{tabular}  & \begin{tabular}[c]{@{}c@{}}basic +\\ small\end{tabular} & \begin{tabular}[c]{@{}c@{}}iso. +\\ large\end{tabular} & barriers & \begin{tabular}[c]{@{}c@{}}intermol.\\ NCIs\end{tabular} & \begin{tabular}[c]{@{}c@{}}intramol.\\ NCIs\end{tabular} & \begin{tabular}[c]{@{}c@{}}all\\ NCIs\end{tabular} & All \\
\midrule
\multirow{5}{*}{PFP} & R2SCAN+D3 & 5.47 & 7.87 & 15.38 & 10.60 & 9.82 & 10.21 & \textbf{9.28} \\
 & R2SCAN & 5.48 & 7.87 & 15.53 & 13.52 & 11.32 & 12.42 & 10.51 \\
 & PBE+D3 & 7.24 & 11.19 & 18.72 & 13.26 & 10.73 & 11.99 & 11.54 \\
 & PBE & 7.21 & 15.28 & 19.12 & 19.69 & 19.15 & 19.42 & 15.43 \\
\midrule
\multirow{5}{*}{DFT} & PBE+D3 & 6.33 & 11.49 & 17.80 & 11.23 & 9.92 & 10.57 & 10.62 \\
& PBE & 6.46 & 15.76 & 16.10 & 17.72 & 19.65 & 18.69 & 14.58 \\
& SCAN+D3 & 5.12 & 6.92 & 14.30 & 9.20 & 6.85 & 8.02 & \textbf{7.94} \\
& SCAN & 5.05 & 8.13 & 13.84 & 11.73 & 8.29 & 10.00 & 8.91 \\
\bottomrule
\end{tabular}
\end{table*}

The evaluation results, summarized in Table~\ref{tab:gmtkn55_results}, highlight the impact of the chosen functional on the accuracy of the PFP model. A direct comparison between the base functionals reveals that PFP-R2SCAN (10.51 kcal/mol) is inherently more accurate than PFP-PBE (15.43 kcal/mol). It should be emphasized that the accuracy of PFP-R2SCAN is better than that of DFT-PBE (14.58 kcal/mol). This improvement demonstrates that the r$^2$SCAN meta-GGA functional provides a more robust and reliable description of diverse chemical interactions, demonstrating the utility of PFP trained with it. Additionally, the accuracy of PFP-R2SCAN is nearly on par with that of meta-GGA-level DFT-SCAN (8.91 kcal/mol).
This advantage is further amplified by the inclusion of Grimme's D3 dispersion correction. We used the correction parameters for r$^2$SCAN as described in \nameref{sec:dft-d3}.  As shown in the results, adding the D3 correction significantly improves the performance of both functionals. The PFP-R2SCAN+D3 achieved a WTMAD-2 score of 9.28 kcal/mol, a notable improvement over the PFP-PBE+D3's score of 11.54 kcal/mol. This underscores that while the D3 correction is crucial for accurately capturing noncovalent interactions, the choice of the underlying functional remains a key determinant of overall performance.
Furthermore, the performance of PFP-R2SCAN+D3 is again highly competitive when compared to conventional DFT-PBE calculations. Notably, our model (9.28 kcal/mol) outperforms the widely-used PBE functional with DFT-D3 correction, which recorded a WTMAD-2 score of 10.62 kcal/mol on the identical subset of data. While meta-GGA functionals with DFT-D3 correction, SCAN+D3, show even higher accuracy (7.94 kcal/mol), our PFP model's ability to surpass the performance of widely used DFT-PBE+D3 underscores its potential as a highly efficient and accurate alternative for chemical simulations. The results confirm that PFP-R2SCAN, particularly when paired with the D3 correction, provides a powerful method that overcomes the inherent limitations of the PBE functional for a universal evaluation of molecular domains.

\paragraph{Surface energies}
To validate the ability of our PFP model for surface domains, we evaluated the surface energy, which is one of the most fundamental properties of surface domains. 
Following the protocol of existing research with DFT~\cite{abhirup2017surface}, we compared the experimental values with the average surface energies calculated on the fcc(100), (110) and (111) surfaces. Each surface energy was calculated using a 30-layer slab, with each layer containing a single atomic layer. The central 10 layers were fixed, while the top and bottom 10 layers underwent surface relaxation until the atomic forces reached less than 0.001 eV/\ang.
The lattice constants were determined using the Birch--Murnaghan equation of state~\cite{PhysRevB.70.224107} and fixed during the surface relaxation.
The surface energy, $\sigma$, was calculated as
\begin{equation}
    \sigma = \frac{E_{\mathrm{slab}} - n \epsilon_{\mathrm{bulk}}}{2A}
\end{equation}
where, $E_{\mathrm{slab}}$ is the energy of the slab structure, $\epsilon_{\mathrm{bulk}}$ is the energy of the bulk structure per atom, $A$ is the area of the surface and $n$ is the number of atoms in the slab structure. While previous studies have employed linear fit method~\cite{Boettger1998linearfit} to mitigate energy convergence errors in DFT calculations, we adopted a simple approach since NNPs including PFP generally exhibit no such convergence errors.

\begin{figure}[t]
    \centering
    \begin{subfigure}[b]{0.48\textwidth}
        \includegraphics[width=\linewidth]{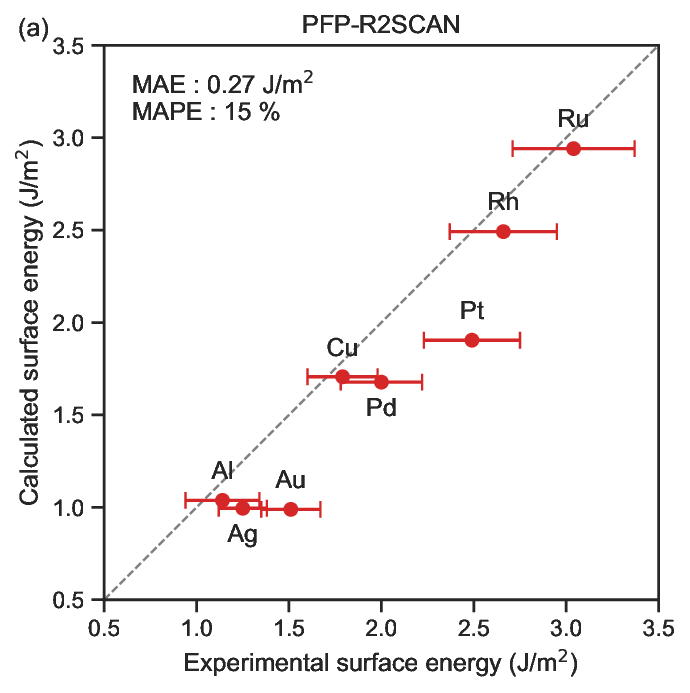}
    \end{subfigure}
    \begin{subfigure}[b]{0.48\textwidth}
        \includegraphics[width=\linewidth]{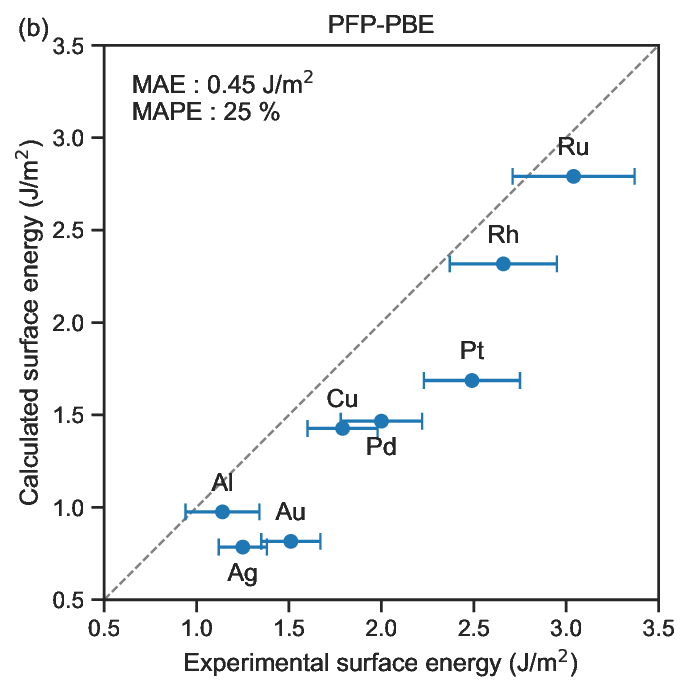}
    \end{subfigure}
    \caption{Reproducibility of the experimental surface energy using (a) PFP-R2SCAN and (b) PFP-PBE. Error bars indicate experimental uncertainties.}
    \label{fig:surf}
\end{figure}

The calculated surface energy results are presented in Figure \ref{fig:surf} and Table \ref{tab:surf}. Following the existing DFT study~\cite{abhirup2017surface}, we performed calculations for Al, Cu, Ru, Rh, Pd, Ag, Pt, and Au, treating all materials as fcc structures—even though Ru is in fact hcp. Note that the slab structures of these elements are included in both PFP-PBE and PFP-R2SCAN datasets.
In Figure \ref{fig:surf}, PFP-PBE significantly underestimates surface energies compared to experimental values, with an MAE of 0.452 $\mathrm{J/m^2}$. The DFT-PBE MAE of 0.434 $\mathrm{J/m^2}$ shows a similar trend of underestimation, confirming the known limitations of the PBE functional. The PBE functional is widely recognized for its inability to accurately capture intermediate-range and long-range vdW interactions~\cite{abhirup2017surface}, which is likely the cause of this underestimation.

\begin{table*}[ht]
    \centering
    \caption{MAE, mean bias error (MBE), and mean absolute percentage error (MAPE) of surface energies against experimental values for multiple PFP calculation modes and DFT functionals}
    \begin{tabular}{llccc}
        \toprule
         Method & calc\_mode or functional & MAE ($\mathrm{J/m^2}$) & MBE ($\mathrm{J/m^2}$) & MAPE (\%)  \\
         \midrule
         \multirow{4}{*}{PFP} & R2SCAN+D3 & 0.21 & -0.15 & 11 \\
                              & R2SCAN & 0.27 & 0.27 & 15 \\
                              & PBE+D3 & 0.29 & -0.27 & 14 \\
                              & PBE & 0.45 & 0.45 & 25 \\
        \midrule
        \multirow{4}{*}{DFT}~\cite{abhirup2017surface}  & SCAN+rVV10 & 0.12 & 0.04 & 6 \\
                              & SCAN & 0.25 & 0.25 & 13 \\
                              & PBE & 0.43 & 0.43 & 24 \\
                              & LDA & 0.15 & -0.04 & 7 \\
         \bottomrule
    \end{tabular}
    \label{tab:surf}
\end{table*}

In Table \ref{tab:surf}, DFT-LDA demonstrates excellent agreement with experimental values, with an MAE of 0.15 $\mathrm{J/m^2}$. This performance is attributed to the accidental cancellation of errors—where the underestimation of exchange energy compensates for the underestimation of correlation energy~\cite{abhirup2017surface}. While LDA produces good results for surface energies, it is generally known to have inferior performance in predicting molecular and crystalline structures compared to GGA functionals like PBE. Therefore, LDA is not suitable for developing uMLIPs aimed at reproducing diverse domains.
In contrast, PFP-R2SCAN achieves surface energy reproduction comparable to that of SCAN functionals—which, as demonstrated in previous subsections, show higher accuracy than PBE for molecules and crystals—while also improving surface energy accuracy. The achieved MAE of 0.27 $\mathrm{J/m^2}$ is nearly identical to the 0.25 $\mathrm{J/m^2}$ MAE obtained in DFT-SCAN. This is likely due to the ability of the r$^2$SCAN functional to capture intermediate-range vdW interactions as a meta-GGA functional, albeit slightly less than the SCAN functional~\cite{acsmaterialsau.2c00059}. The experimental uncertainty in surface energies is typically around 0.2 $\mathrm{J/m^2}$, and it should be noted that for half of the elements shown in Figure \ref{fig:surf}, the surface energies calculated using PFP-R2SCAN fall within the experimental error range.
 Long-range vdW interactions were addressed through D3 corrections, resulting in improved MAE values for both PFP-PBE+D3 and PFP-R2SCAN+D3 compared to their respective base versions. While PFP-PBE+D3 demonstrated reproduction accuracy comparable to PFP-R2SCAN, PFP-R2SCAN+D3 achieved even higher accuracy with an MAE of 0.21 $\mathrm{J/m^2}$. Although there is a slight discrepancy with the MAE values obtained using vdW density functional, SCAN+rVV10~\cite{PhysRevX.6.0410}, these results still represent good accuracy considering the experimental uncertainty in surface energies is typically around 0.2 $\mathrm{J/m^2}$.

\subsection{Bridging simulation and reality: melting points calculation}
\begin{figure}
    \centering
    \centering
    \includegraphics[width=\linewidth]{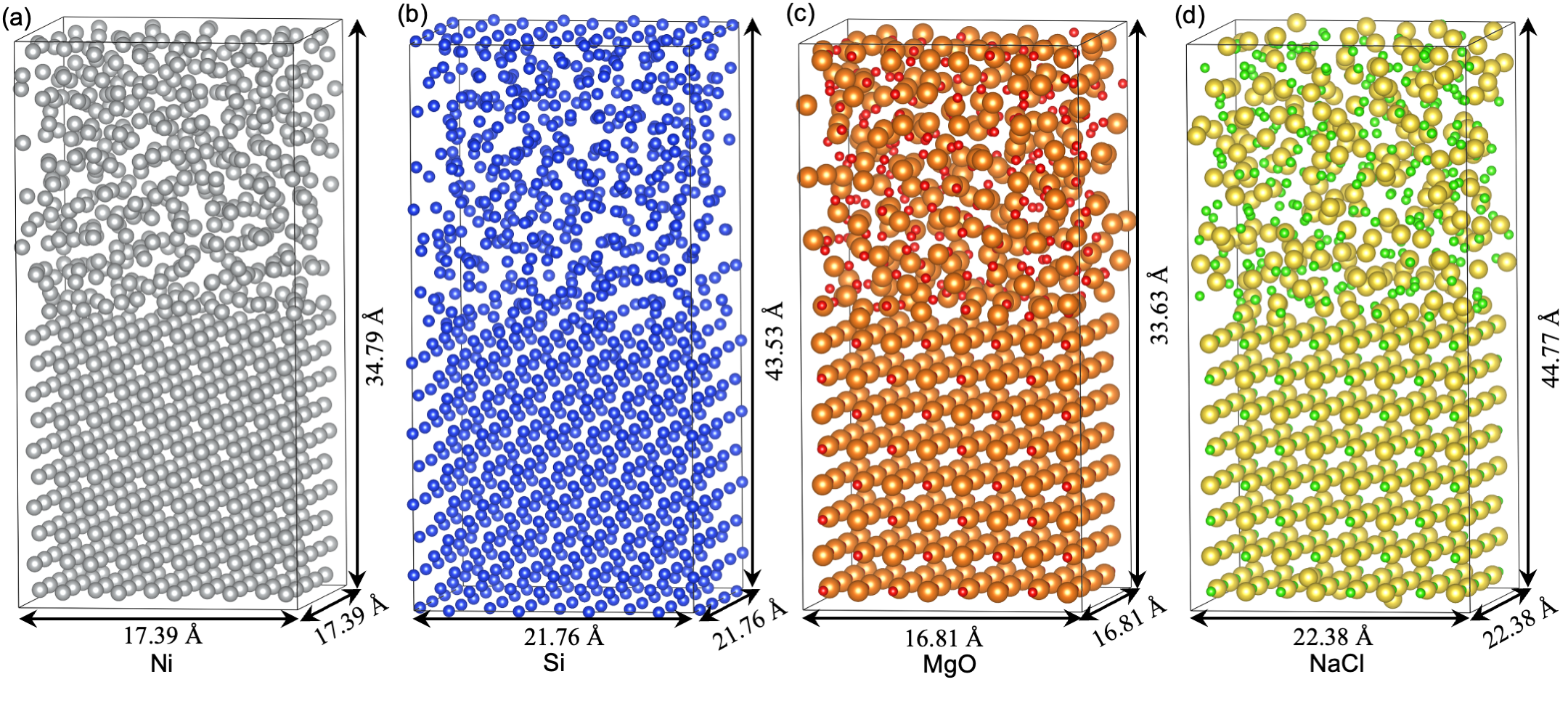}
    \caption{Examples of solid-liquid interface structures, visualized by VESTA\cite{momma2011vesta}, for melting-point determination of (a) metals, (b) covalent solids, (c) oxides, and (d) ionic compounds.}
    \label{fig:mpstc-short}
\end{figure}

\begin{figure}
    \centering
    \centering
    \begin{subfigure}[b]{0.48\textwidth}
        \includegraphics[width=\linewidth]{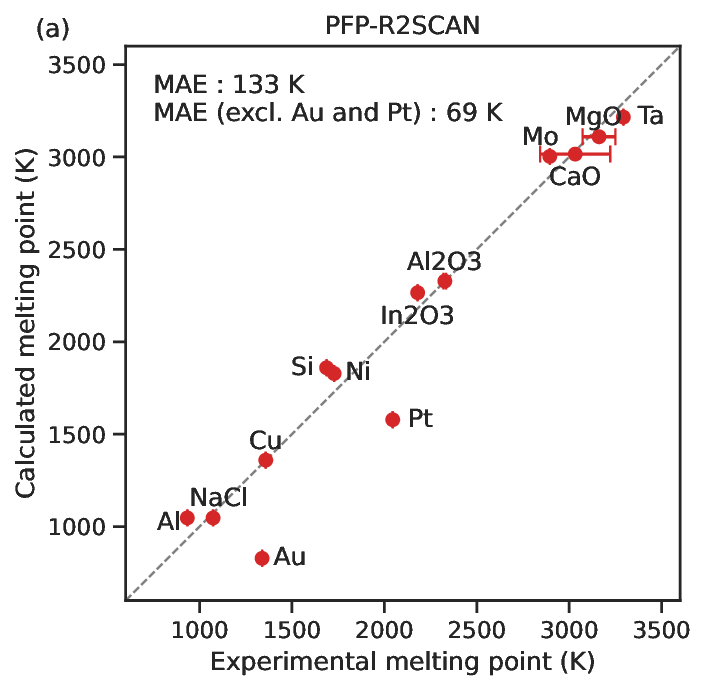}
    \end{subfigure}
    \begin{subfigure}[b]{0.48\textwidth}
        \includegraphics[width=\linewidth]{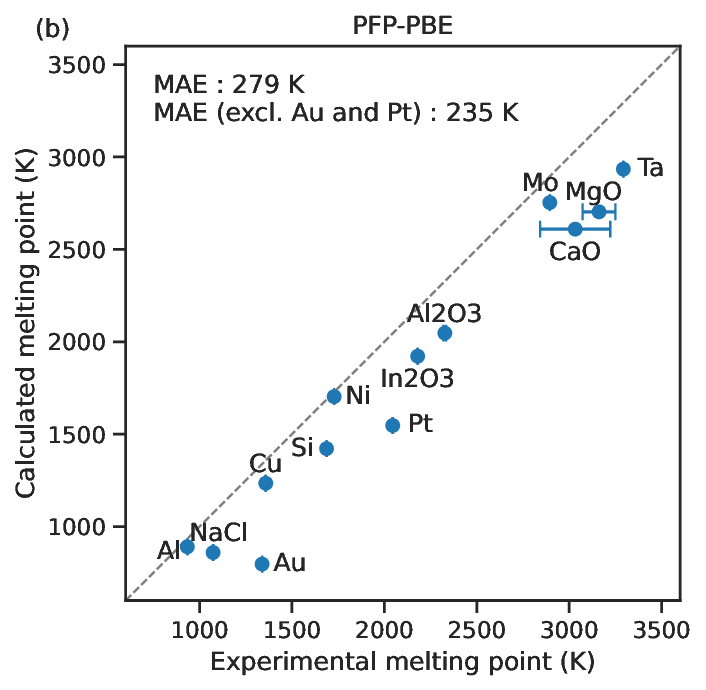}
    \end{subfigure}
    \caption{Comparison of melting temperatures determined by (a) PFP-R2SCAN and (b) PFP-PBE with experimental values. The MAEs including and excluding the data points for \ce{Au} and \ce{Pt} are shown due to underfitting of PFP to the two materials. Details are summarized in Table~\ref{tab:exp_melting_points} and Table~\ref{tab:calc_melting_points}.}
    \label{fig:mpvsexp}
\end{figure}

To validate our PFP models for predicting melting points, we evaluate them on a set of representative materials, including oxides (\ce{MgO}, \ce{CaO}, \ce{Al2O3}, and \ce{In2O3}), an ionic compound (\ce{NaCl}), a covalent solid (\ce{Si}), and several metals (\ce{Ni}, \ce{Cu}, \ce{Al}, \ce{Au}, \ce{Pt}, \ce{Mo}, and \ce{Ta}). The two-phase coexistence method\cite{vega2008determination} is used to determine melting points. To limit finite-size errors to approximately 25 K, all atomistic models comprise roughly 1,000 atoms\cite{andolina2024predicting,remsing2018refined,dorner2018melting} (Fig.~\ref{fig:mpstc-short}, Fig.~\ref{fig:mpstc-1}). Details of the method and simulation setups are summarized in the Supporting Information.
The predicted melting points are in excellent agreement with DFT and experimental values reported in the literature\cite{lee2022ab,jinnouchi2019fly,andolina2024predicting,cazorla2007ab,dorner2018melting,rang2019first,shah2023first,pozzo2013melting}, except for \ce{Au} and \ce{Pt}. The details are summarized in Table~\ref{tab:calc_melting_points}.

We first compare melting points between PFP-PBE and PFP-R2SCAN. For PFP-PBE, the MAE with respect to experimental values is 279 K while the MAE for PFP-R2SCAN is 133 K (Fig.~\ref{fig:mpvsexp}). 
Thus, the melting-point determination is improved through the use of PFP-R2SCAN. 
Specifically, predictions for the ionic material and the oxides are significantly improved, while the accuracy for metals is preserved when transitioning from PFP-PBE to PFP-R2SCAN. For the covalent solid, \ce{Si}, PFP-R2SCAN gives a melting point of 1859 K, significantly higher than the 1421 K predicted by PFP-PBE. The increase in the melting point results from stabilization of the covalent phase upon moving from the PBE to the SCAN functional\cite{remsing2018refined}.

Two outliers are the melting points of the solids, \ce{Au} and \ce{Pt}. Both PFP-PBE and PFP-R2SCAN yield melting points of about 830 K and 1570 K, respectively, which is significantly lower than the experimental value\cite{Kittel2005} of 1338 K and 2045 K, respectively. 
The causes of the underestimation are different for \ce{Au} and \ce{Pt}. For \ce{Au}, previous studies reported a melting point of 1181 K by means of the local density approximation\cite{weck2020determination} (LDA) functional and 972 K by means of MLIPs fitting to PBE functional\cite{andolina2024predicting}. This indicates that LDA/PBE inherently underestimates the melting point of \ce{Au}. This intrinsic deficiency persists in the r$^2$SCAN functional; recent assessments show that r$^2$SCAN predicts a bulk modulus for \ce{Au} (153.5 GPa) that remains significantly lower than the experimental value (182.0 GPa)\cite{liu2024assessing}. This ``unphysical softness'' of the lattice at 0 K renders the crystal susceptible to premature thermal instability, a failure that can likely only be corrected by including nonlocal many-body effects as captured by the Random Phase Approximation (RPA)\cite{grabowski2015random}.  Furthermore, the PFP shows an underfitting behavior for \ce{Au} because the value from PFP-PBE is lower than the reported PBE value\cite{andolina2024predicting}.  
Thus, the underestimation of the melting point of \ce{Au} comes from both the approximation of XC functional and underfitting of PFP to the training data. 
In contrast, an \textit{ab initio} study\cite{belonoshko2012high} reproduced the experimental melting point of \ce{Pt} by GGA. In this case, the underestimation arises solely from the underfitting of PFP to the \ce{Pt} metal.

\section{Conclusion}

In this study, we developed PFP v8, a uMLIP based on the meta-GGA level r$^2$SCAN functional, with the aim of bridging the gap between reality and simulation by improving the accuracy vs. experimental values.

Comprehensive validation confirmed that PFP-R2SCAN significantly outperforms PBE-based DFT calculations and MLIPs across diverse domains, including crystals, molecules, and surfaces. For crystal structures, PFP-R2SCAN reduced the MAE to 80 meV/atom---two-thirds of the error observed in PBE functionals using Hubbard $U$ or empirical corrections---without requiring any empirical corrections itself. For molecules, we demonstrated improved accuracy for various properties, such as reaction barriers and intermolecular interactions, through the GMTKN55 benchmark based on high-accuracy calculations, which are considered sufficiently close to experimental values. Regarding surface structures, PFP-R2SCAN addressed the significant underestimation characteristic of PBE functionals; furthermore, when combined with the D3 correction (PFP-R2SCAN+D3), it achieved an MAE of 0.210 J/m², a value comparable to experimental measurement error. Notably, these results were produced without fine-tuning for specific domains and under identical computational conditions, with the exception of the D3 correction. This demonstrates that PFP v8, as a uMLIP not limited to specific domains like crystals, enables high reproducibility of reality via the r$^2$SCAN functional, unaffected by inconsistencies arising from differing computational conditions.

One of the key results is that the accuracy vs. experimental values for the melting points of multiple materials, calculated via long-duration MD simulations, was substantially improved from an MAE of 279 K with PFP-PBE to 133 K with PFP-R2SCAN. Performing such extensive MD simulations is practically impossible with standard DFT calculations, and conventional PBE-based MLIPs have historically shown large deviations from experimental values. This indicates that PFP v8 enables simulations that are much closer to reality---an achievement unattainable by conventional methods---by combining the computational speed of MLIPs with the high accuracy of the r$^2$SCAN functional.

We anticipate that PFP v8, a uMLIP capable of handling diverse domains without fine-tuning while maintaining high accuracy vs. experimental values, will be utilized in various materials exploration and research efforts. Furthermore, we believe that future uMLIP development should focus on enhancing reproducibility of experimental values and bridging the gap between reality and simulation, rather than solely reproducing results fitting the specific computational conditions used for dataset construction.

\begin{acknowledgement}

PFP v8 is developed using our in-house supercomputer and AI Bridge Cloud Infrastructure (ABCI) 2.0 and 3.0 provided by Japan’s National Institute of Advanced Industrial Science and Technology (AIST) and AIST Solutions Co., Ltd. The use of ABCI 3.0 is supported by the ABCI 3.0 Development Acceleration Program.
We thank ENEOS Holdings, Inc., Matlantis Corp., Matlantis Inc., and some Matlantis users for their collaboration in the evaluation of early versions of PFP v8 and for providing valuable feedback that contributed to its development. Additionally, the authors acknowledge a report from a Matlantis user regarding the melting 
point of Au.
The authors thank Ryo Nagai and Iori Kurata for the valuable discussion on DFT calculations and XC functionals. The authors thank Nontawat Charoenphakdee for his critical reading of the manuscript and valuable comments.

\end{acknowledgement}

\begin{suppinfo}
See supporting information for DFT calculation details and simulation details.
\end{suppinfo}

\bibliography{main}

\makeatletter\@input{supplement_aux.tex}\makeatother
\end{document}


\tableofcontents
\clearpage
\section{Details of DFT Calculation}
\label{si:dft_details}

\begin{figure}[t]
    \centering
    \begin{minipage}[t]{\textwidth}
         \centering
         \includegraphics[width=\textwidth]{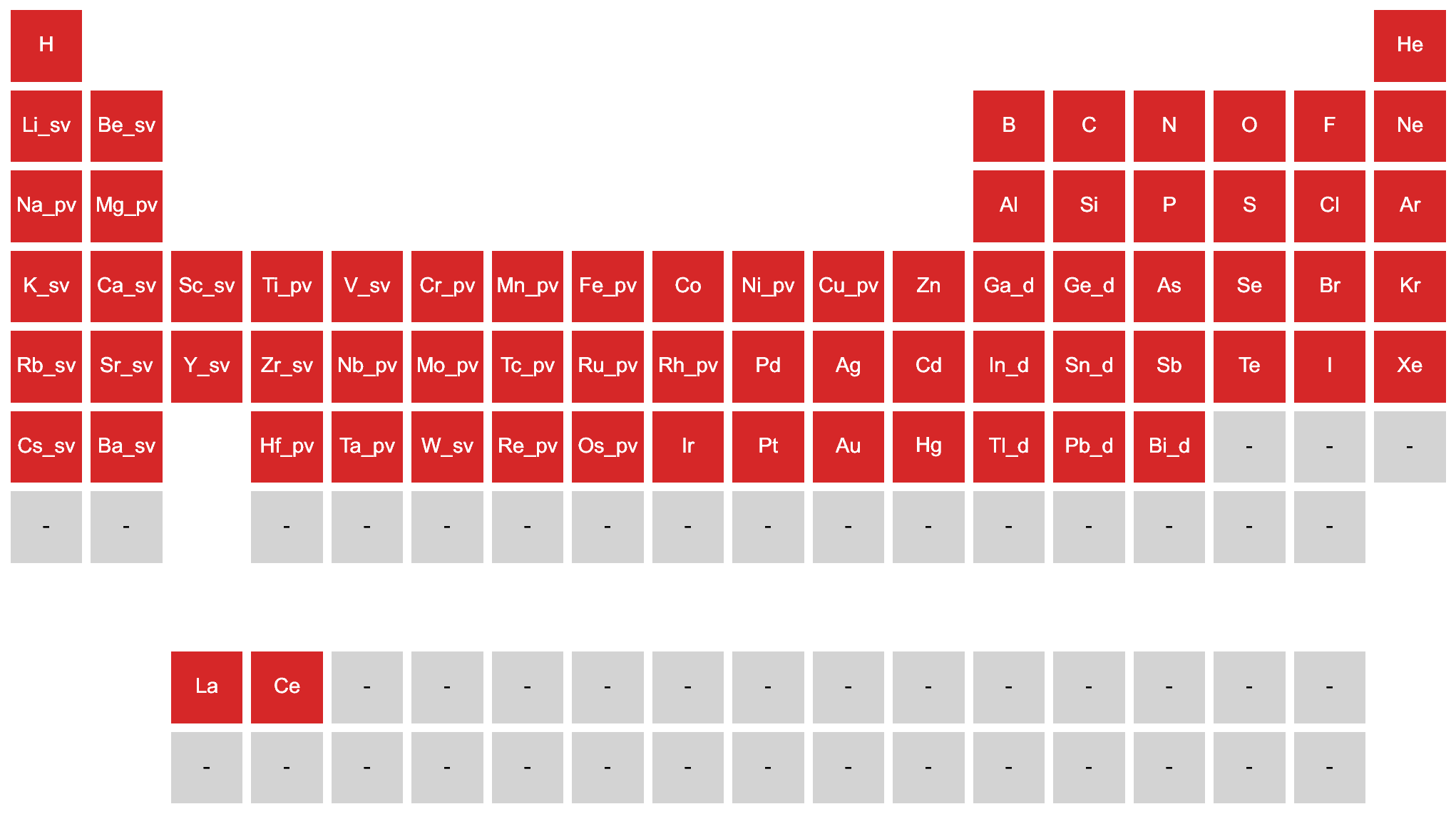}
         \subcaption{PFP-R2SCAN with PBE\_64 set}
    \end{minipage}
    \begin{minipage}[t]{\textwidth}
        \centering
        \includegraphics[width=\textwidth]{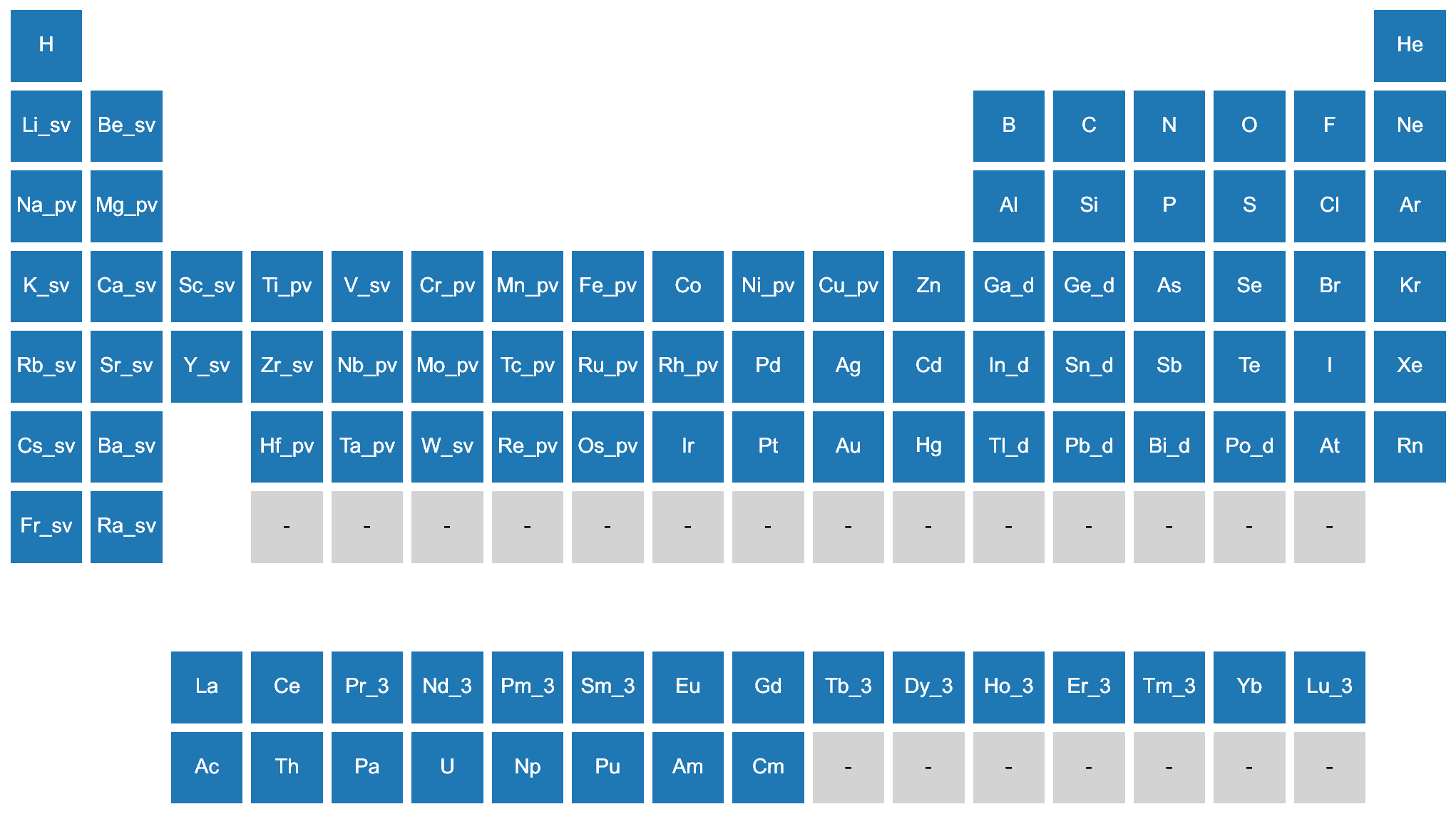}
        \subcaption{PFP-PBE/PBE+U with PBE\_54 set}
    \end{minipage}
    \caption{Pseudopotential set for (a) PFP-R2SCAN with PBE\_64 set provided by VASP and (b) PFP-PBE/PBE+U with PBE\_54 set. The periodic table symbols, such as ``H'', ``Li\_sv'' and ``Na\_pv'', indicate the names of the pseudopotentials.}
    \label{fig:vasp_pp_set}
\end{figure}

The specific computational conditions and INCAR tags employed in our calculations are summarized in Table~\ref{tab:vasp_parameters} and Figure~\ref{fig:vasp_pp_set}.

\begin{table*}[h]
    \caption{Our VASP calculation conditions for r$^2$SCAN and PBE/PBE+U, along with comparisons with MatPES and PBE MPRelax calculations}
    \label{tab:vasp_parameters}
    \centering
    \resizebox{\textwidth}{!}{
    \begin{tabular}{lrrrr}
    \toprule
    & PFP-R2SCAN & PFP-PBE/PBE+U~\cite{takamoto2022towards} & MatPES~\cite{matpes} & PBE MPRelax~\cite{jain_commentary_2013} \\
    \midrule
    INCAR tags & & & & \\
    \midrule
    ALGO & All & Normal & Normal  & Fast \\
    EDIFF & 1e-6 & 1e-4 & 1e-5 &  5e-5 * $N_{atoms} $\\
    ENAUG & 1360 & - &  1360 & -\\
    ENCUT & 680 &  520 & 680 & 520 \\
    ISMEAR & 0 & 0 & 0 &  -5 \\
    ISPIN & 2 & 2 &  2 & 2 \\
    LASPH & .TRUE. & .FALSE. & .TRUE. & .TRUE. \\
    LMAXMIX & 6 & 2, 4, or 6 & 6 & 2, 4, or 6 \\
    LMIXTAU & .TRUE. & .FALSE. & .TRUE. & .FALSE. \\
    
    LREAL & .FALSE. & Auto & .FALSE. & Auto \\
    
    PREC & Accurate & Accurate &  Accurate &  Accurate \\
    SIGMA & 0.05 & 0.05 & 0.05 & 0.05 \\
    \midrule
    other settings & & & & \\
    \midrule
    Pseudopotential & PBE\_64 & PBE\_54 & PBE\_64 & PBE\_54 \\
    k-point sampling & KSPACING=0.22  & 1000 kppa & KSPACING=0.22 &1000 kppa \\
    \bottomrule
    \end{tabular}
    }
\end{table*}{}

\clearpage

\section{Details of molecular benchmarks on GMTKN55}
\label{si:gmtkn55}

\begin{table*}[h]
    \centering
    \caption{GMTKN55 Validation Targets and Subsets}
    \label{tab:validation_targets_en}
    \begin{tabular}{|l|p{4cm}|p{6cm}|}
        \hline
        \textbf{Target Name} & \textbf{Validation Target} & \textbf{Subsets} \\
        \hline
        basic + small & Basic properties and reaction energies of small systems & W4-11, ALKBDE10, YBDE18, AL2X6, HEAVYSB11, NBPRC, ALK8, G2RC, BH76RC, FH51, TAUT15, DC13 \\
        \hline
        iso. + large & Reaction energies of large systems and isomerization reactions & MB16-43, DARC, RSE43, BSR36, CDIE20, ISO34, ISOL24, C60ISO \\
        \hline
        barriers & Reaction barrier heights & BH76, BHPERI, BHDIV10, INV24, BHROT27, PX13, WCPT18 \\
        \hline
        intermolecular NCIs & Intermolecular non-covalent interactions & RG18, ADIM6, S22, S66, HEAVY28, WATER27, CARBHB12, PNICO23, HAL59, IL16 \\
        \hline
        intramolecular NCIs & Intramolecular non-covalent interactions & IDISP, ICONF, ACONF, Amino20x4, PCONF21, MCONF, SCONF, BUT14DIOL \\
        \hline
        all NCIs & Intermolecular and intramolecular interactions & Sub-benchmark datasets for intermolecular NCIs and intramolecular NCIs \\
        \hline
        All & All 38 targets in GMTKN55 & All of the above datasets \\
        \hline
    \end{tabular}
\end{table*}

\newpage
\section{Details of surface energy calculations}
\label{si:surface_energies}
\begin{figure}[h]
    \centering
    \includegraphics[width=\linewidth]{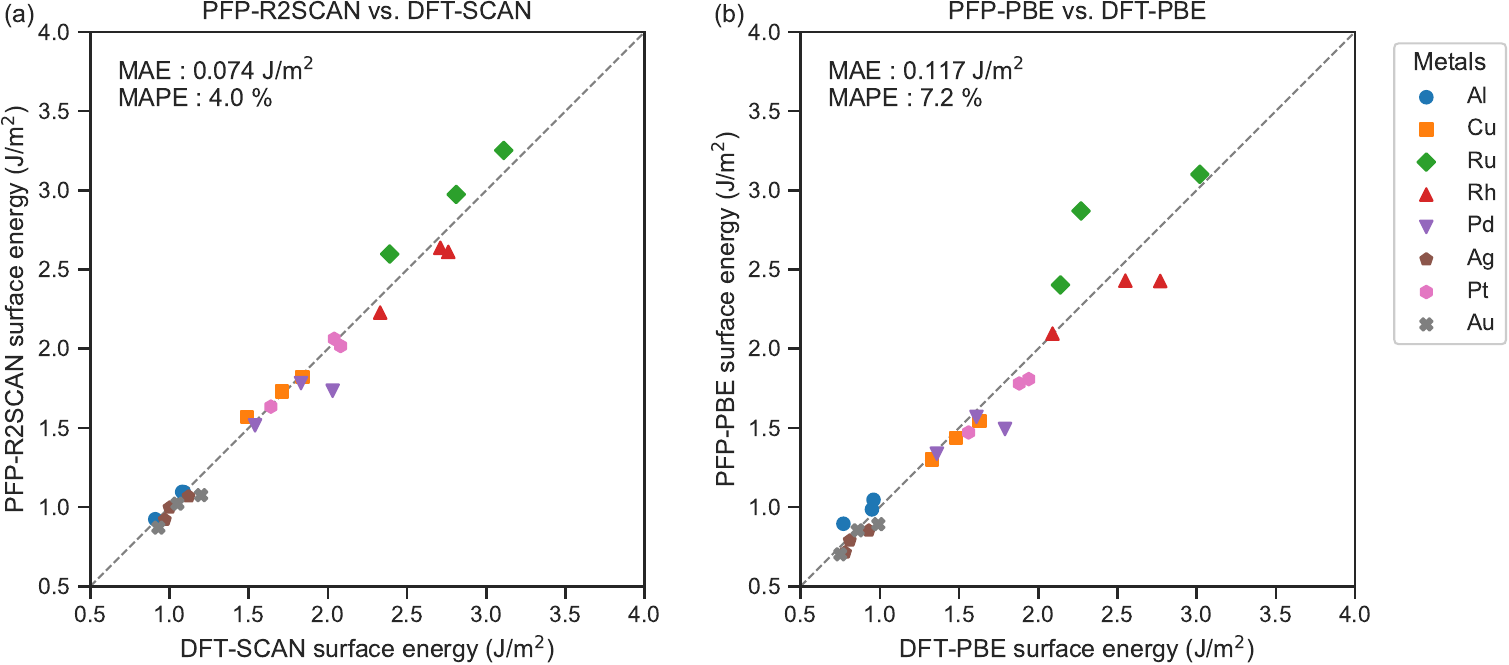}
    \caption{Comparison of surface energies calculated using PFP with reference DFT results\cite{abhirup2017surface} for the fcc (100), (110), and (111) surfaces of Al, Cu, Ru, Rh, Pd, Ag, Pt, and Au. (a) PFP-R2SCAN versus DFT-SCAN; (b) PFP-PBE versus DFT-PBE. Each symbol type represents a specific metallic element, and multiple data points per element correspond to different crystallographic surfaces evaluated.}
    \label{fig:surf_e_vs_dft}
\end{figure}

Figure \ref{fig:surf_e_vs_dft} compares the surface energies calculated using PFP with reference DFT results from DFT-SCAN calculations for the fcc crystal planes (100), (110), and (111) of Al, Cu, Ru, Rh, Pd, Ag, Pt, and Au.
The MAE of PFP-R2SCAN relative to DFT-SCAN is 0.074 $\mathrm{J/m^2}$, while that of PFP-PBE relative to DFT-PBE is smaller than 0.117 $\mathrm{J/m^2}$.

Despite comparing results with different exchange-correlation functionals (r$^2$SCAN versus SCAN), the MAE of PFP-R2SCAN remains smaller than that of PFP-PBE when comparing results using the same exchange-correlation functional. This superior DFT reproduction capability in PFP-R2SCAN can be attributed to the more stringent computational parameters employed in the PFP-R2SCAN dataset.
As mentioned in the main text, the PFP-R2SCAN dataset uses a vacuum layer thickness of 20 \ang, larger than the 10 \ang~used in the PFP-PBE dataset. Additionally, the $k$-point sampling method in the PFP-R2SCAN dataset is more suitable for slab structures.

\newpage
\section{Details of melting points calculation}
\label{si:mp}
\subsection{Experimental references}
\begin{table*}[h]
    \centering
\begin{tabular}{lclSSc}
\toprule
Material & MP ID & Space group & {Exp. $T_\mathrm{m}$ (K)} & {Exp. err. (K)} & Reference \\
\midrule
CaO & mp-2605 & $Fm\bar{3}m$  & 3032.5 & 189.5 & \citenum{liang2018complete} \\
MgO & mp-1265 & $Fm\bar{3}m$  & 3161.5 & 88.5 & \citenum{liang2018complete} \\
NaCl & mp-22862 & $Fm\bar{3}m$ & 1073.0 & 1.0 &  \citenum{yamada1993melting}\\
Si & mp-149 & $Fd\bar{3}m$  & 1687.0 & {-} & \citenum{Kittel2005} \\
Al & mp-134 & $Fm\bar{3}m$ & 933.5 & {-} & \citenum{Kittel2005} \\
Cu & mp-30 & $Fm\bar{3}m$  & 1358 & {-} & \citenum{Kittel2005} \\
Ni & mp-23 & $Fm\bar{3}m$  & 1728 & {-} & \citenum{Kittel2005} \\
\ce{In2O3} & mp-2255 & $Ia\bar{3}$  & 2183.0 & 10 & \citenum{Schneider1961} \\
\ce{Al2O3} & mp-1143 & $R\bar{3}c$ & 2327.0 & 7 &  \citenum{qiu2018simulation} \\
Au & mp-81 & $Fm\bar{3}m$  & 1338 & {-} & \citenum{Kittel2005} \\
Mo & mp-129 & $Im\bar{3}m$  & 2895 & {-} & \citenum{Kittel2005} \\
Ta & mp-5 & $Im\bar{3}m$  & 3293 & {-} & \citenum{Kittel2005} \\
Pt & mp-126 & $Fm\bar{3}m$  & 2045 & {-} & \citenum{Kittel2005} \\
\bottomrule
\end{tabular}
    \caption{Experimental melting points ($T_\mathrm{m}$) and reported uncertainties for materials. The corresponding bulk crystal structures used in melting-point determination were retrieved from the Materials Project\cite{jain_commentary_2013} database (v2025.09.25), specifically selecting experimentally stable structures. The Materials Project (MP) IDs are listed.}
    \label{tab:exp_melting_points}
\end{table*}

\subsection{Method}
The coexistence method\cite{vega2008determination} is used to determine melting points.
To automate the process of melting-point determination, we adopt the algorithm proposed by Zhu et al.\cite{zhu2021fully}. In this algorithm, initial guesses, $T_\mathrm{high}^{i=1}$ and $T_\mathrm{low}^{i=1}$, are provided for the upper and lower bounds of the temperature window at the first iteration, where the superscript $i$ indicates the iteration index. We then perform NPT simulations independently at the upper and lower bound temperatures and determine whether each state is solid or liquid. The temperature window is updated as follows: If both states are solid, then $T_\mathrm{low}^{i} = T_\mathrm{high}^{i-1} $ and $T_\mathrm{high}^{i} = T_\mathrm{high}^{i-1} + T_\mathrm{high}^{i-1} - T_\mathrm{low}^{i-1} $; If the material is solid at $T_\mathrm{low}^{i-1}$ and liquid at $T_\mathrm{high}^{i-1}$, $T_\mathrm{low}^{i} = T_\mathrm{low}^{i-1} + (T_\mathrm{high}^{i-1} - T_\mathrm{low}^{i-1})/2.0$; If the material is liquid at both temperatures,  $T_\mathrm{high}^{i} = T_\mathrm{low}^{i-1} $ and $T_\mathrm{low}^{i} = T_\mathrm{low}^{i-1} - (T_\mathrm{high}^{i-1} - T_\mathrm{low}^{i-1})/2.0 $. The rule is summarized in Eq.~\ref{eq:update}. 

\begin{equation}
\begin{cases}
\text{Both solid:} & 
\begin{cases}
T_\mathrm{low}^{i} = T_\mathrm{high}^{i-1}, \\[0.5em]
T_\mathrm{high}^{i} = T_\mathrm{high}^{i-1} + (T_\mathrm{high}^{i-1} - T_\mathrm{low}^{i-1})
\end{cases} \\[1.2em]

\text{Solid--liquid:} &
\begin{cases}
T_\mathrm{high}^{i} = T_\mathrm{high}^{i-1},\\[0.5em] 
T_\mathrm{low}^{i} = T_\mathrm{low}^{i-1} 
+ \dfrac{T_\mathrm{high}^{i-1} - T_\mathrm{low}^{i-1}}{2}
\end{cases} \\[1.2em]

\text{Both liquid:} &
\begin{cases}
T_\mathrm{high}^{i} = T_\mathrm{low}^{i-1}, \\[0.5em]
T_\mathrm{low}^{i} = T_\mathrm{low}^{i-1} 
- \dfrac{T_\mathrm{high}^{i-1} - T_\mathrm{low}^{i-1}}{2}
\end{cases}
\end{cases}
\label{eq:update}
\end{equation}

To check whether a material is in a solid or liquid phase, we use order parameters (OP), $q_{\mathrm{oct}}$, $q_{\mathrm{tet}}$, $q_{\mathrm{bcc}}$, which are proposed by \citet{zimmermann2017assessing}, and $q_{6}$ by \citet{steinhardt1983bond} 
and implemented in Pymatgen\cite{ong2013python}. 
It is important to note that the choice of order parameters is not limited to the mentioned order parameters; rather, it depends on the local atomic environment.
The cutoff parameters in the order parameters determine which atoms are considered neighbors of the center atoms. A proper value should be larger than the position of the first peak in a radial distribution function. 
We first calculate the local order parameters of all atoms in a structure. If an atom's order parameter is larger than a predefined threshold ($\theta_{\text{solid}}$), the atom is considered to be in a solid environment; otherwise, it is considered to be in a liquid environment. In addition, if an atom has fewer than $N_\text{min}$ neighbors, its order parameter is set to zero, indicating a liquid-like state. Lastly, we measure the percentage of atoms in a solid environment. If this percentage exceeds a predefined threshold ($\theta_{\%\text{solid}}$), the structure is considered to be in a solid phase; otherwise, it is considered to be in a liquid phase. All relevant parameters are summarized in Table~\ref{tab:mpsimparam}.

Next, the procedure for constructing solid–liquid interface structures and performing coexistence simulations is described. All the geometry optimizations and MD simulations are performed by the open-source code atomic simulation environment (ASE). 
Solid-liquid interface structures are created as follows. The unit cell of each material is retrieved from the Materials Project\cite{jain_commentary_2013} and then is subjected to geometric optimization with FrechetCellFilter, allowing both atomic positions and cell parameters to relax. The optimized unit cell is replicated in accordance with the supercell size provided by the user. While half of the supercell along the z axis is fixed, the other half of the supercell is randomized by performing an NVE simulation at an initial temperature of 20,000 K. The timestep of the simulation is set to 2.0 fs and the number of MD steps is 5,000. Then, the melted half is slightly cooled down at a temperature of 4000 K by carrying out an MD simulation in the NVT ensemble with 1,000 MD steps and 2.0 fs timestep. The temperature is controlled using a Langevin thermostat with a friction coefficient of 0.002 in ASE internal units.
Next, all atoms in the supercell are allowed to equilibrate by an NPT simulation at a testing temperature defined by the user and zero external pressure. The temperature is kept by the Nos\'{e}-Hoover thermostat with a time constant of 20 fs. The pressure is maintained by the Parrinello-Rahman barostat with a pfactor of $1.0\times10^{6} \space \text{GPa}\cdot\text{fs}^{2}$. The last snapshot of the MD is taken as the initial structure for the production run. 

The coexistence of a liquid-solid interface is tested by performing an NPT simulation at a testing temperature defined by the user and a pressure of 101,325 Pascal. The setups of the thermostat and barostat are the same as above. All timesteps are set to 1.0 fs.
\begin{figure*}
    \centering
    \includegraphics[width=\linewidth]{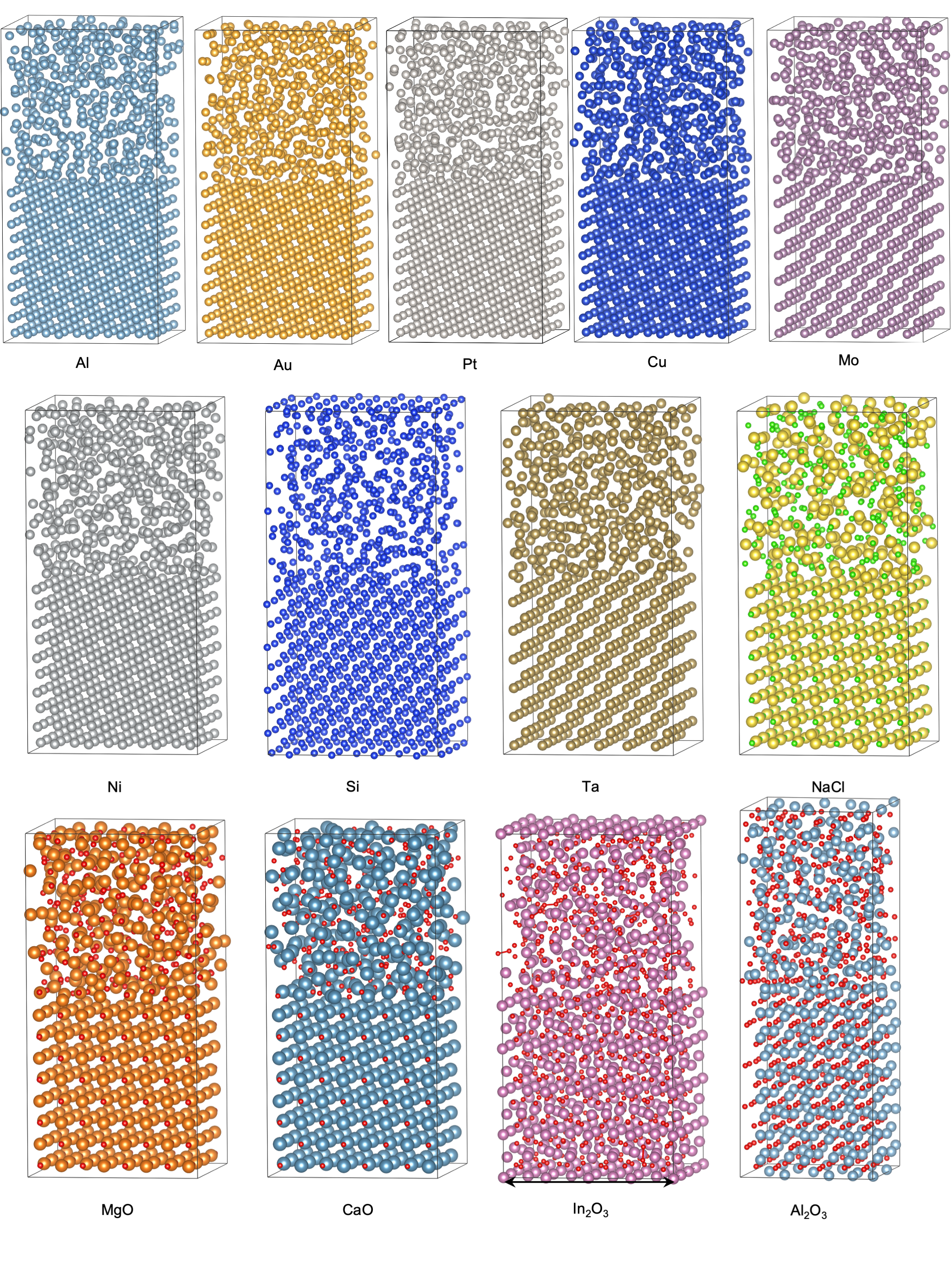}
    \caption{Solid-liquid interface structures from PFP-R2SCAN calc\_mode, visualized by VESTA\cite{momma2011vesta}, for melting-point determination. Cell parameters are compiled in Table.~\ref{tab:cellparam}.}
    \label{fig:mpstc-1}
\end{figure*}

\begin{table}[htbp]
\centering
\caption{Supercell parameters (in \AA) for different interface systems using calc\_mode PFP-R2SCAN and PFP-PBE. $N_\text{tot}$ denotes the total number of atoms.}
\begin{tabular}{l|c|ccc|ccc}
\hline
 System & $N_\text{tot}$ & \multicolumn{3}{c|}{R2SCAN} & \multicolumn{3}{c}{PBE } \\
 &  & $a$(\AA) & $b$(\AA) & $c$ (\AA)& $a$(\AA) & $b$(\AA) & $c$(\AA) \\
\hline
Al          & 1000 & 19.98 & 19.98 & 39.96 & 20.22 & 20.22 & 40.45 \\
Al$_2$O$_3$ & 1080 & 16.46 & 14.26 & 38.97 & 16.64 & 14.41 & 39.34 \\
Au          & 1000 & 20.63 & 20.63 & 41.26 & 20.78 & 20.78 & 41.56 \\
CaO         & 1024 & 19.24 & 19.24 & 38.49 & 19.37 & 19.37 & 38.75 \\
Cu          & 1000 & 17.88 & 17.88 & 35.76 & 18.10 & 18.10 & 36.20 \\
In$_2$O$_3$ & 1280 & 20.37 & 20.37 & 40.74 & 20.61 & 20.61 & 41.22 \\
MgO         & 1024 & 16.81 & 16.81 & 33.63 & 17.03 & 17.03 & 34.07 \\
Mo          & 864  & 18.91 & 18.91 & 37.83 & 18.95 & 18.95 & 37.90 \\
NaCl        & 1024 & 22.38 & 22.38 & 44.77 & 22.79 & 22.79 & 45.59 \\
Ni          & 1000 & 17.39 & 17.39 & 34.79 & 17.55 & 17.55 & 35.10 \\
Pt          & 1000 & 19.71 & 19.71 & 39.43 & 19.85 & 19.85 & 39.70 \\
Si          & 1024 & 21.76 & 21.76 & 43.53 & 21.86 & 21.86 & 43.72 \\
Ta          & 864  & 19.85 & 19.85 & 39.70 & 19.90 & 19.90 & 39.81 \\
\hline
\end{tabular}
\label{tab:cellparam}
\end{table}

\begin{table*}[h!]
\centering
\resizebox{\textwidth}{!}{
\begin{tabular}{l c c c c c c S S S}
\hline
Material & Supercell & MD steps & OP Type &
$\theta_{\text{solid}}$ &
$\theta_{\%\text{solid}}$ &
Cutoff (\ang) &
$N_\text{min}$ &
{$T_\mathrm{high}^{i=1}$ (K)}&
{$T_\mathrm{low}^{i=1}$ (K)}\\
\hline
CaO      & $4\times4\times8$     & 100000 & $q_{\text{oct}}$ & 0.25 & 0.50  & 2.90  & {--} & 4000 & 2000 \\
MgO      & $4\times4\times8$     & 100000 & $q_{\text{oct}}$ & 0.25 & 0.50  & 2.50  & {--} & 3500 & 2500 \\
NaCl     & $4\times4\times8$     & 100000 & $q_{\text{oct}}$ & 0.25 & 0.50  & 3.20  & {--} & 1500 & 500 \\
Si       & $4\times4\times8$     & 100000 & $q_{\text{tet}}$ & 0.25 & 0.50  & 2.75 & {--} & 2000 & 1000 \\
Al       & $5\times5\times10$    & 100000 & $q_6$            & 0.25 & 0.50  & 3.25 & 10 & 1500 & 500 \\
Cu       & $5\times5\times10$    & 100000 & $q_6$            & 0.25 & 0.65 & 3.00 & 10 & 2000 & 1000 \\
Ni       & $5\times5\times10$    & 100000 & $q_6$            & 0.25 & 0.55 & 2.85 & 10 & 2000 & 1000 \\
Al$_2$O$_3$ & $2\times3\times3$  & 125000 & $q_{\text{oct}}$ & 0.25 & 0.50  & 2.50  & 4  & 3000 & 2000 \\
In$_2$O$_3$ & $2\times2\times4$  & 100000 & $q_{\text{oct}}$ & 0.25 & 0.50  & 2.60 & 4  & 2500 & 1500 \\
Mo       & $6\times6\times12$    & 100000 & $q_{\text{bcc}}$ & 0.25 & 0.20  & 3.05 & 4  & 3300 & 2300 \\
Ta       & $6\times6\times12$    & 100000 & $q_{\text{bcc}}$ & 0.25 & 0.20  & 3.16 & 4  & 3700 & 2700 \\
Au       & $5\times5\times10$    & 100000 & $q_6$            & 0.25 & 0.50  & 3.35 & 10 & 1500 & 500 \\
Pt       & $5\times5\times10$    & 125000 & $q_6$            & 0.25 & 0.50 & 3.18 & 10 & 2500 & 1500 \\
\hline
\end{tabular}
}
\caption{Simulation parameters for melting point determination}
\label{tab:mpsimparam}
\end{table*}

\begin{table*}[]
\centering
\begin{tabular}{l*{6}{S}}
\toprule
\multirow{2}{*}{Material} & 
\multicolumn{2}{c}{PFP  $T_\mathrm{m}$  (K)} & 
\multicolumn{2}{c}{MLIP $T_\mathrm{m}$ (K)} &
\multicolumn{2}{c}{DFT  $T_\mathrm{m}$  (K)} \\
\cmidrule(lr){2-3} \cmidrule(lr){4-5} \cmidrule(lr){6-7}
 & {R2SCAN} & {PBE} & {SCAN} & {PBE} & {SCAN} & {PBE} \\
\midrule
CaO   & 3015 & 2609 & 3057\textsuperscript{\emph{a}}  & 2659\textsuperscript{\emph{a}} &  &  \\
MgO  & 3109 & 2703 & 3181\textsuperscript{\emph{a}}  & 2786\textsuperscript{\emph{a}}  & 3032\textsuperscript{\emph{f}}  & 2747\textsuperscript{\emph{f}} \\
NaCl & 1046 & 859 &  & 875\textsuperscript{\emph{g}} & &\\
Si  & 1859 & 1421 & 1825\textsuperscript{\emph{b}}  &  1431\textsuperscript{\emph{b}} & 1842\textsuperscript{\emph{e}} & 1449\textsuperscript{\emph{e}} \\
Al & 1046 & 890 &  981\textsuperscript{\emph{b}}  & 837\textsuperscript{\emph{b}} & & \\
Cu  & 1359 & 1234 &  & 1196\textsuperscript{\emph{c}}  & &\\
Ni & 1828 & 1703 &  & 1580\textsuperscript{\emph{c}} & & 1637\textsuperscript{\emph{h}} \\
\ce{In2O3} &  2265 & 1921 &  & & &\\
\ce{Al2O3} &  2328 & 2046 &  & & &\\
Au &  828 & 796 &  & 972\textsuperscript{\emph{c}} & & \\
Mo &  3003 & 2753 &  & 2573\textsuperscript{\emph{c}} & &2894\textsuperscript{\emph{d}} \\
Ta & 3215 & 2934 &  & & & 3270\textsuperscript{\emph{i}} \\
Pt & 1578 & 1546 &  & & &  \\
\bottomrule
\end{tabular}
\caption{Calculated melting points ($T_\mathrm{m}$) from PFP-R2SCAN and PFP-PBE, compared with available DFT literature values.}
\label{tab:calc_melting_points}

\raggedright
\textsuperscript{\emph{a}} Data from Ref.~\citenum{lee2022ab}. Method: coexistence\\
\textsuperscript{\emph{b}} Data from Ref.~\citenum{jinnouchi2019fly}. Method: interface pinning \\
\textsuperscript{\emph{c}} Data from Ref.~\citenum{andolina2024predicting}. Method: coexistence \\
\textsuperscript{\emph{d}} Data from Ref.~\citenum{cazorla2007ab}. Method: coexistence \\
\textsuperscript{\emph{e}} Data from Ref.~\citenum{dorner2018melting}. Method: thermodynamic integration \\
\textsuperscript{\emph{f}} Data from Ref.~\citenum{rang2019first}. Method: thermodynamic integration \\
\textsuperscript{\emph{g}} Data from Ref.~\citenum{shah2023first}. Method: coexistence \\
\textsuperscript{\emph{h}} Data from Ref.~\citenum{pozzo2013melting}. Method: coexistence \\
\textsuperscript{\emph{i}} Data from Ref.~\citenum{taioli2007melting}. Method: coexistence \\
\end{table*}

\subsection{Convergence of temperature windows}
\begin{figure*}
    \centering
    \includegraphics[width=\linewidth]{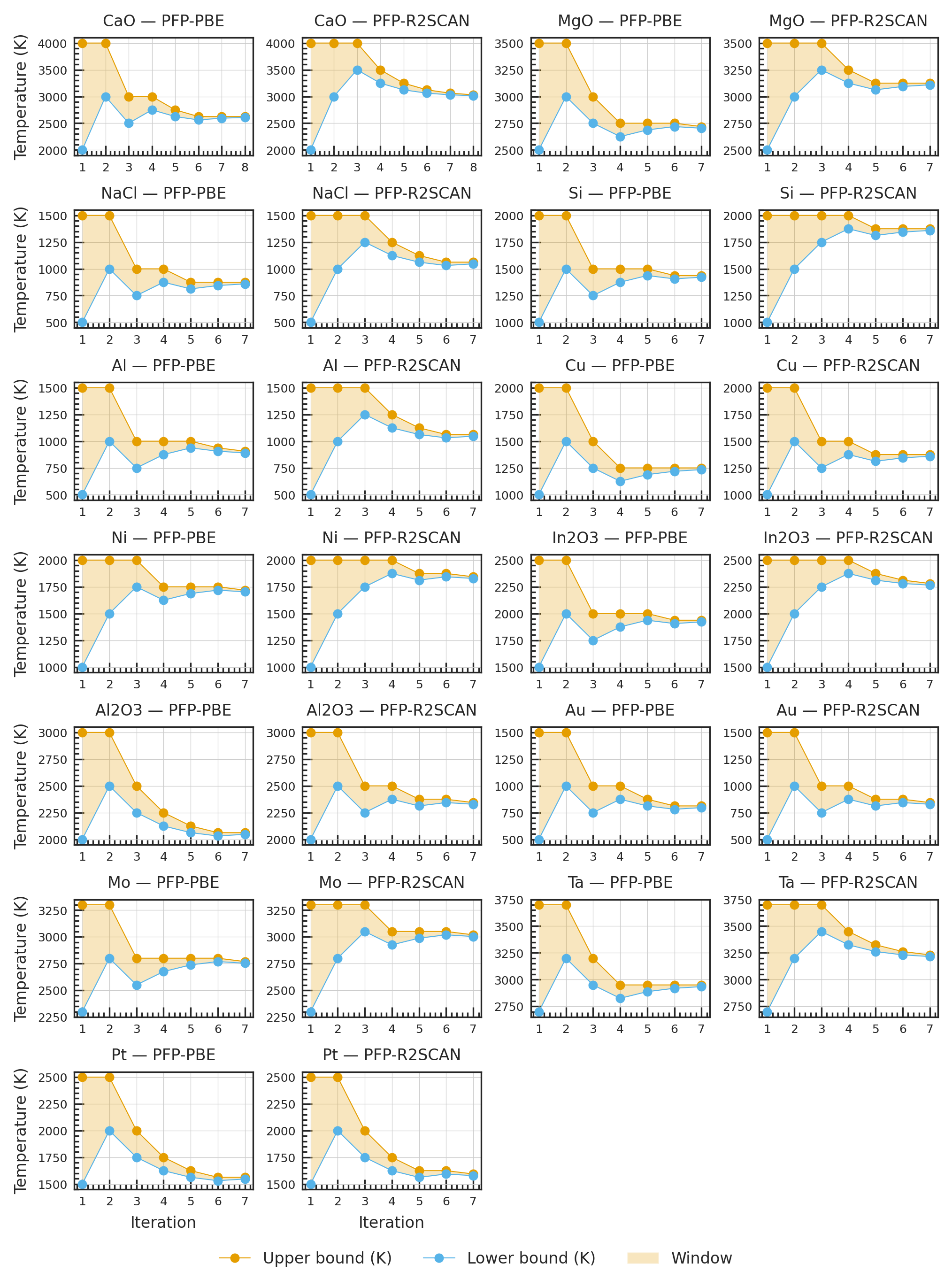}
    \caption{Convergence of the temperature window for materials with PBE and R2SCAN calc\_mode. }
    \label{fig:mpconv}
\end{figure*}
In Figure~\ref{fig:mpconv}, all temperature windows for the three materials converge within 8 iterations because each iteration halves the window width by definition. This approach also has one drawback: it may fail to find the correct temperature if the initial temperature window is far from the computed melting temperature, because the window width decays as $(T_\mathrm{high}^{0}-T_\mathrm{low}^{0}) \times 1/2^i$.

\clearpage
\bibliography{main}